\documentclass[twoside,twocolumn,9pt]{article}
\usepackage{extsizes}
\usepackage[super,sort&compress,comma]{natbib} 
\usepackage[version=3]{mhchem}
\usepackage[left=1.5cm, right=1.5cm, top=1.785cm, bottom=2.0cm]{geometry}
\usepackage{balance}
\usepackage{times,mathptmx}
\usepackage{sectsty}
\usepackage{graphicx} 
\usepackage{lastpage}
\usepackage[format=plain,justification=justified,singlelinecheck=false,font={stretch=1.125,small,sf},labelfont=bf,labelsep=space]{caption}
\usepackage{float}
\usepackage{fancyhdr}
\usepackage{fnpos}
\usepackage[english]{babel}
\addto{\captionsenglish}{%
  
}
\usepackage{array}
\usepackage{droidsans}
\usepackage{charter}
\usepackage[T1]{fontenc}
\usepackage[usenames,dvipsnames]{xcolor}
\usepackage{setspace}
\usepackage[compact]{titlesec}
\usepackage{hyperref}
\usepackage{gensymb}

\usepackage{epstopdf}

\definecolor{cream}{RGB}{222,217,201}

\begin{document}

\pagestyle{fancy}
\thispagestyle{plain}
\fancypagestyle{plain}{

\fancyhead[C]{\includegraphics[width=18.5cm]{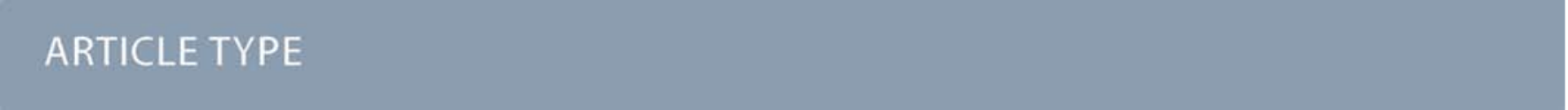}}
\fancyhead[L]{\hspace{0cm}\vspace{1.5cm}\includegraphics[height=30pt]{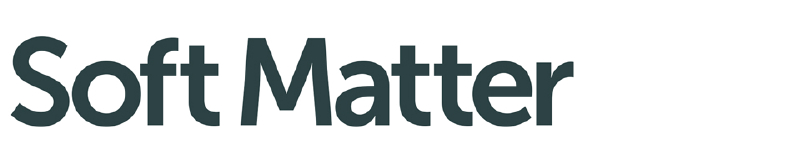}}
\fancyhead[R]{\hspace{0cm}\vspace{1.7cm}\includegraphics[height=55pt]{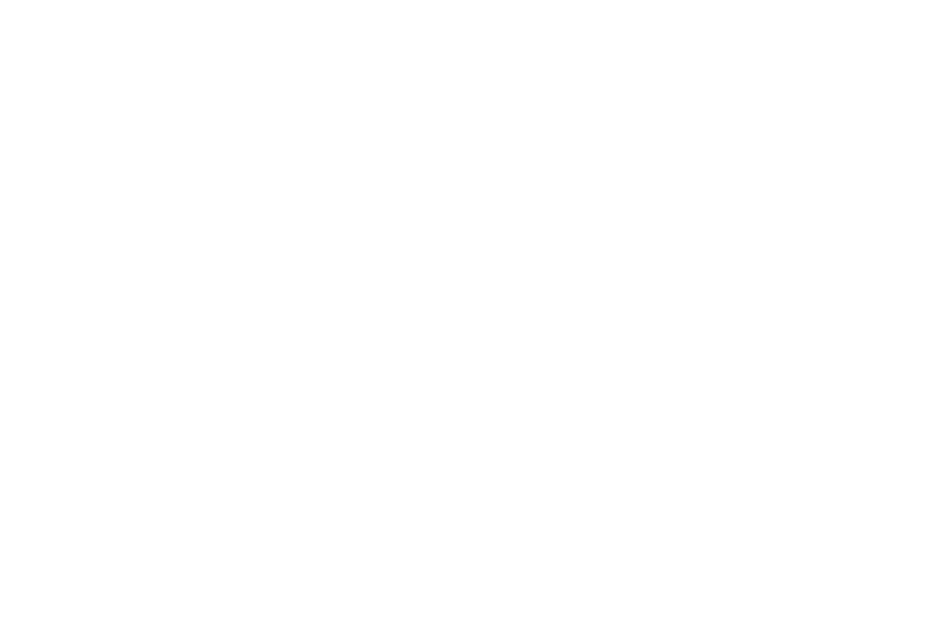}}
\renewcommand{\headrulewidth}{0pt}
}

\makeFNbottom
\makeatletter
\renewcommand\LARGE{\@setfontsize\LARGE{15pt}{17}}
\renewcommand\Large{\@setfontsize\Large{12pt}{14}}
\renewcommand\large{\@setfontsize\large{10pt}{12}}
\renewcommand\footnotesize{\@setfontsize\footnotesize{7pt}{10}}
\makeatother

\renewcommand{\thefootnote}{\fnsymbol{footnote}}
\renewcommand\footnoterule{\vspace*{1pt}%
\color{cream}\hrule width 3.5in height 0.4pt \color{black}\vspace*{5pt}} 
\setcounter{secnumdepth}{5}

\makeatletter 
\renewcommand\@biblabel[1]{#1}            
\renewcommand\@makefntext[1]%
{\noindent\makebox[0pt][r]{\@thefnmark\,}#1}
\makeatother 
\renewcommand{\figurename}{\small{Fig.}~}
\sectionfont{\sffamily\Large}
\subsectionfont{\normalsize}
\subsubsectionfont{\bf}
\setstretch{1.125} 
\setlength{\skip\footins}{0.8cm}
\setlength{\footnotesep}{0.25cm}
\setlength{\jot}{10pt}
\titlespacing*{\section}{0pt}{4pt}{4pt}
\titlespacing*{\subsection}{0pt}{15pt}{1pt}

\fancyfoot{}
\fancyfoot[LO,RE]{\vspace{-7.1pt}\includegraphics[height=9pt]{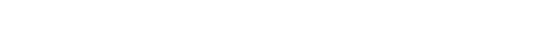}}
\fancyfoot[CO]{\vspace{-7.1pt}\hspace{13.2cm}\includegraphics{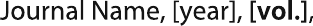}}
\fancyfoot[CE]{\vspace{-7.2pt}\hspace{-14.2cm}\includegraphics{head_foot/RF}}
\fancyfoot[RO]{\footnotesize{\sffamily{1--\pageref{LastPage} ~\textbar  \hspace{2pt}\thepage}}}
\fancyfoot[LE]{\footnotesize{\sffamily{\thepage~\textbar\hspace{3.45cm} 1--\pageref{LastPage}}}}
\fancyhead{}
\renewcommand{\headrulewidth}{0pt} 
\renewcommand{\footrulewidth}{0pt}
\setlength{\arrayrulewidth}{1pt}
\setlength{\columnsep}{6.5mm}
\setlength\bibsep{1pt}

\makeatletter 
\newlength{\figrulesep} 
\setlength{\figrulesep}{0.5\textfloatsep} 

\newcommand{\topfigrule}{\vspace*{-1pt}%
\noindent{\color{cream}\rule[-\figrulesep]{\columnwidth}{1.5pt}} }

\newcommand{\botfigrule}{\vspace*{-2pt}%
\noindent{\color{cream}\rule[\figrulesep]{\columnwidth}{1.5pt}} }

\newcommand{\dblfigrule}{\vspace*{-1pt}%
\noindent{\color{cream}\rule[-\figrulesep]{\textwidth}{1.5pt}} }

\makeatother

\twocolumn[
  \begin{@twocolumnfalse}
\vspace{3cm}
\sffamily
\begin{tabular}{m{4.5cm} p{13.5cm} }

\includegraphics{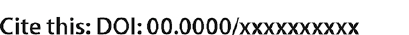} & \noindent\LARGE{\textbf{In-vivo characterization of optically trapped Brownian probes 'at a glance'}} \\
\vspace{0.3cm} & \vspace{0.3cm} \\

 & \noindent\large{R. Vaippully, S. R. Vaibavi, S. Bajpai, and B. Roy} \\

\includegraphics{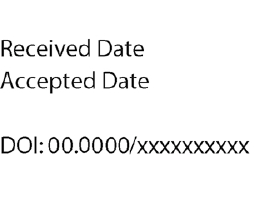} & \noindent\normalsize{Calibration of optically trapped particles in-vivo has been complicated given the frequency dependence and spatial inhomogeneity of the cytoplasmic viscosity, and the requirement of accurate knowledge of the medium refractive index. Further, it has been demonstrated that the medium viscosity is dependent upon the measurement probe leading to reliability issues for measurements with even micrometer sized particles. Here, we employ a recent extension of Jeffery's model of viscoelasticity in the microscopic domain to fit the passive motional power spectra of micrometer-sized optically trapped particles embedded in a viscoelastic medium. We find excellent agreement between the 0 Hz viscosity in MCF7 cells and the typical values of viscosity in literature, between 2 to 16 mPa sec expected for the typical concentration of proteins inside the cytoplasmic solvent. This bypasses the dependence on probe size by relying upon small thermal displacements. Our measurements of the relaxation time also match values reported with magnetic tweezers, at about 0.1 sec. Finally, we calibrate the optical tweezers and demonstrate the efficacy of the technique to the study of in-vivo translational motion.} \\

\end{tabular}

 \end{@twocolumnfalse} \vspace{0.6cm}

  ]

\renewcommand*\rmdefault{bch}\normalfont\upshape
\rmfamily
\section*{}
\vspace{-1cm}






\section{Statement of significance}

Our manuscript shows how to calibrate the thermal motion of a probe trapped in-vivo using optical tweezers. We use a new theory that makes the viscosity frequency dependent quite akin to the Jeffery's model. We find that the fits to the power spectral density are very good and we can extract the 0 Hz (DC) viscosity that match well with literature. We can make the optical tweezers calibration "at a glance". In view of the fact that calibration of optical tweezers in-vivo has remained challenging due to the variability of the intracellular viscosity and refractive index, this fit to the Jeffery's model solves the problem completely.

\section{Introduction}
Microrheology is the study of flow of matter at micron length scales and microliter sample volumes. It is particularly important when rare or precious materials are involved, like in bio-physical studies. These can be performed in an environment where conventional rheological tools cannot reach inside the  cells\cite{fabry,allan}. Rheological characterization of cells is relevant for early diagnosis of diseases like malaria \cite{malaria} and migration of cancerous cells \cite{cancer1}.  The Red Blood Cells (RBC) in malaria infected patients are known to have different elasticity than healthy ones. The Extra-Cellular Matrix (ECM) of a cell in the human body interacts and changes the stiffness of the cytoskeleton in healthy cells via mechanical adaptation, the exact mechanism of which is not very well understood. A better match of the rheological features of the ECM and the cell is required for their adherence and cell mobility. Normally, when the ECM is too stiff and the cell cannot change its cytoskeletal features, there is no adherence. In the case of cancer cells, this mechanism is suppressed thereby allowing attachment inspite of stiffness mismatch and executes metastasis. Thus, one promising strategy to address cancer could be to understand and revive this cellular mechanism to match rheological features. In view of these facets, in-vivo rheology is useful. Further, intracellular viscoelasticity has a role in diffusion of molecules and performance of chemical reactions, not to mention, such rheology also enables calibration of externally applied forces and torques on the system. 
 
 Recently, the cell cytoplasm was shown to be a poroelastic medium, \cite{poroelastic} implying a medium with an elastic meshlike network into which a viscoelastic medium is suspended. There have been numerous attempts to ascertain both the viscosity\cite{puchkov,kuimova} and the viscoelasticity\cite{berret} of the cell cytoplasm. However, it is well known that the viscosity results are prone to the size of the probe used \cite{sizeeff}. Any probe larger than 100 nm shall yield a viscosity larger than the native one, when probed actively. It is here, that a passive detection technique that moves the probe by small amounts can be expected to avoid the elastic mesh and provide information about the cytoplasmic fluid. Thus we study the passive thermal fluctuations of a probe using optical tweezers. Attempts to use optical tweezers in-vivo \cite{oddershede,roop,goldman,reihani} have been complicated due to the problems in quantifying the optical trap stiffness accurately \cite{gross}, mainly due to the variable nature of the intracellular refractive index and the viscosity. We use a new theory to directly ascertain the unknown parameters from the fit  "at a glance". 
 
 It is generally believed that the viscoelasticity of a medium automatically implies a segment-wise power-law behavior of the complex elastic modulus ($G^*(\omega)$ = A $\omega^{\beta}$) in the frequency domain \cite{powerlaw}.
 Recently, there has been an attempt to establish the macroscopic viscoelasticity of the medium from the microscopic Stokes Oldroyd-B model that showed a frequency-dependent viscosity quite akin to the Jeffery's model \cite{shuvojit}. We use this new strategy to attack the problem of intra-cellular viscoelasticity and find that the motional power spectral density for 
 spherical polystyrene particles of radius $ a_0 = 0.5 \mu m$ fits well to the suggested model. We extract the 0 Hz (DC) viscosity and find a good match with the established intra-cellular viscosity values. We also find the relaxation time of the intracellular medium and the optical trap stiffness also emerge from the fitting parameters.

\section{Theory}

The frequency- dependent viscosity in an incompressible low Reynolds number viscoelastic medium comprising of a solvent and a polymer solute dissolved in it has been found to be given by the following expression derived from the Stokes Oldroyd-B model for linear microscopic viscoelasticity \cite{shuvojit}
\begin{equation}
     \mu(\omega)=\mu_s + \frac{\mu_p}{-i\omega\lambda+1}  
    \label{oldroyd}
\end{equation}
Where $\mu_s$ is the zero frequency solvent viscosity, $\mu_p$ is the zero frequency polymer viscosity and $\lambda$ is the polymer relaxation time. This expression is very similar to the Jeffery's model of frequency dependent viscosity with the coefficients labelled differently.  Thus the viscosity of the solution at zero frequency would be, $\mu_0$=$\mu_s+\mu_p$.
Solving for the power spectral density of an optically trapped particle in a viscoelastic fluid, we get \cite{shuvojit}
\begin{equation}
 <x(\omega)x^*(\omega)>= \frac{2k_BT}{\gamma_0}\frac{(\frac{(1+\frac{\mu_p}{\mu_s})}{\lambda^2}+\omega^2)}{[(\frac{\kappa}{\gamma_0\lambda}-\omega^2)^2+\omega^2(\frac{\kappa}{\gamma_0}+\frac{1}{\lambda}(1+\frac{\mu_p}{\mu_s}))^2]}
 \label{PSD}
\end{equation}
The term $\kappa$ signifies the trap stiffness and $\gamma_0$ is the drag coefficient for only the solvent. The exact expression for the $\gamma_0$ is given by,

\begin{equation}
\gamma_0=6\pi\mu_s a_0
\label{solvent}
\end{equation}

In order to correlate with experimentally obtained power spectral density curves of translational motion, we rewrite the eq. \ref{PSD}, as 

\begin{equation}
 <x(\omega)x^*(\omega)>= \beta^2 A\frac{(\frac{(1+\frac{\mu_p}{\mu_s})}{\lambda^2}+\omega^2)}{[(\frac{\kappa}{\gamma_0\lambda}-\omega^2)^2+\omega^2(\frac{\kappa}{\gamma_0}+\frac{1}{\lambda}(1+\frac{\mu_p}{\mu_s}))^2]}
 \label{fit}
\end{equation}

where the A coefficient indicates the amplitude in terms of Volts$^2$/Hz and the calibration factor is $\beta$ in (m/Volt) quite akin to the conventional calibration factor for normal media \cite{erik}. The calibration factor for the translational signal is related to temperature as 

\begin{equation}
 \beta^2 A = \frac{2 k_B T}{\gamma_0}
 \label{fit1}
\end{equation}
Thus, the calibration factor $\beta$ is given as 

\begin{equation}
 \beta = \sqrt{\frac{2 k_B T}{A \gamma_0}}
 \label{fit2}
\end{equation}
Fitting the power spectral density with this equation we can extract the values of relative viscosity($\frac{\mu_s+\mu_p}{\mu_s}$) of solution and polymer relaxation time constant($\lambda$).

\section{Experimental details}

\begin{figure}[h]
\centering
  \includegraphics[width=\linewidth]{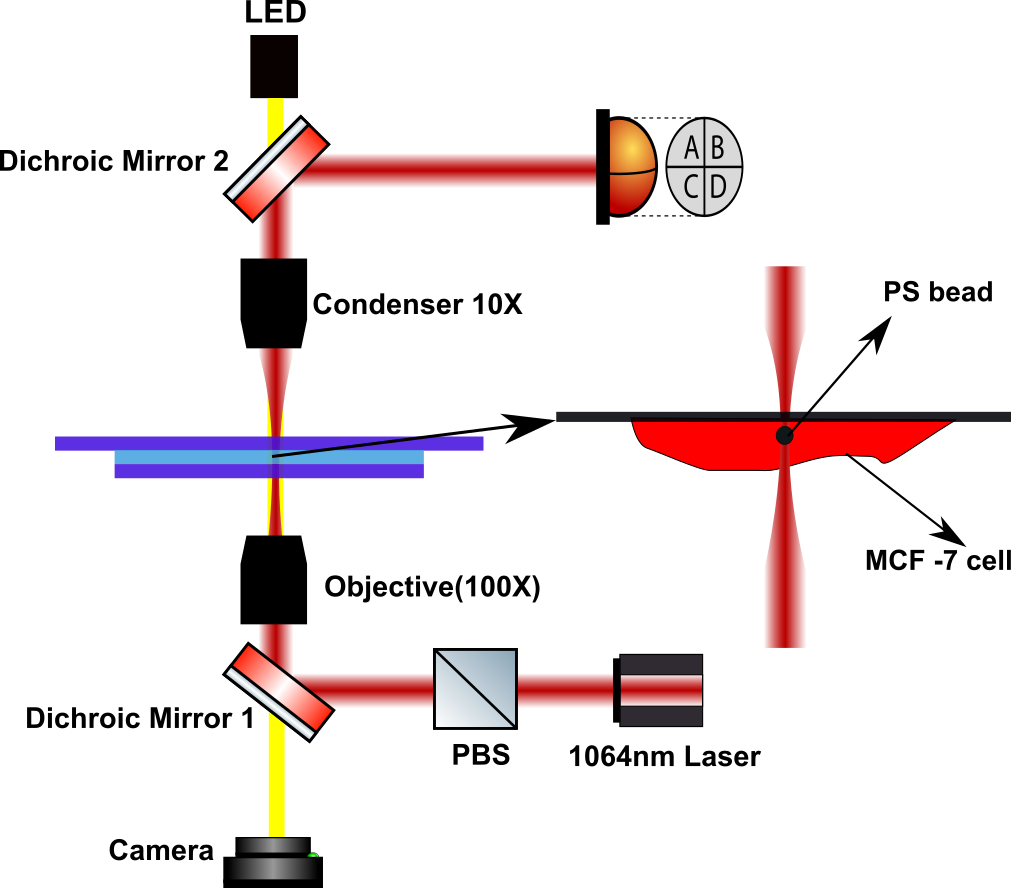}
  \caption{The schematic of the experimental set up. The MCF7 cell is attached to the top surface of a sample chamber with the probe particle located inside the cell. The Optical tweezers light traps the particle and then the forward scattered light is used to perform translational measurements.}
  \label{schematic}
\end{figure}

To perform the experiment, we place a polystyrene particle (1  $\mu$m diameter) inside a cell attached to the glass slide (Blue Star, 75 mm length, 25 mm width and a thickness of 1.1 mm) of a sample chamber assembled, as shown in Fig. \ref{schematic}. The other side of the sample chamber is formed by a cover slip (Blue Star, number 1 size, english glass). The cell is close to the top surface of the sample chamber and illuminated in an inverted microscopy configuration using the Optical Tweezers kit (OTKB-M, Thorlabs USA) \cite{rahul}. The illumination objective is a 1.3 NA, 100x oil immersion objective from Olympus at the bottom with the illumination aperture being overfilled. The collection objective at the top is  E Plan 10x, 0.25 N.A. air-immersion objective from Nikon. The laser used for optical trapping is a diode laser from Lasever at 1064 nm wavelength which typically has a maximum power of 1.7 Watt with about 400 mW power in the sample plane. However, for the experiments performed in this manuscript, the laser was set to 200 mW power at the sample plane. An LED lamp illuminates the sample, coupled in using a dichroic mirror, as shown in Fig. \ref{schematic}. Another dichroic mirror couples the broadband visible light out of the path of the laser and illuminates a CMOS camera (Thorlabs, USA). 

The translational motion is recorded by impinging the forward scattered light onto a Quadrant Photodiode (QPD) (Thorlabs, USA). This has four quadrants indicated by A, B, C, and D in Fig. \ref{schematic}. The top half minus the bottom half yields y displacement while the right half minus the left half yields the x displacement \cite{basudev}. The bandwidth of this detector is 40 KHz. The position signals emerging from the QPD are acquired by the computer using data acquisition cards (National Instruments, NI PCI 6143) which has a bandwidth of 40 KHz. 

\begin{figure}
 \centering
 \includegraphics[width=\linewidth]{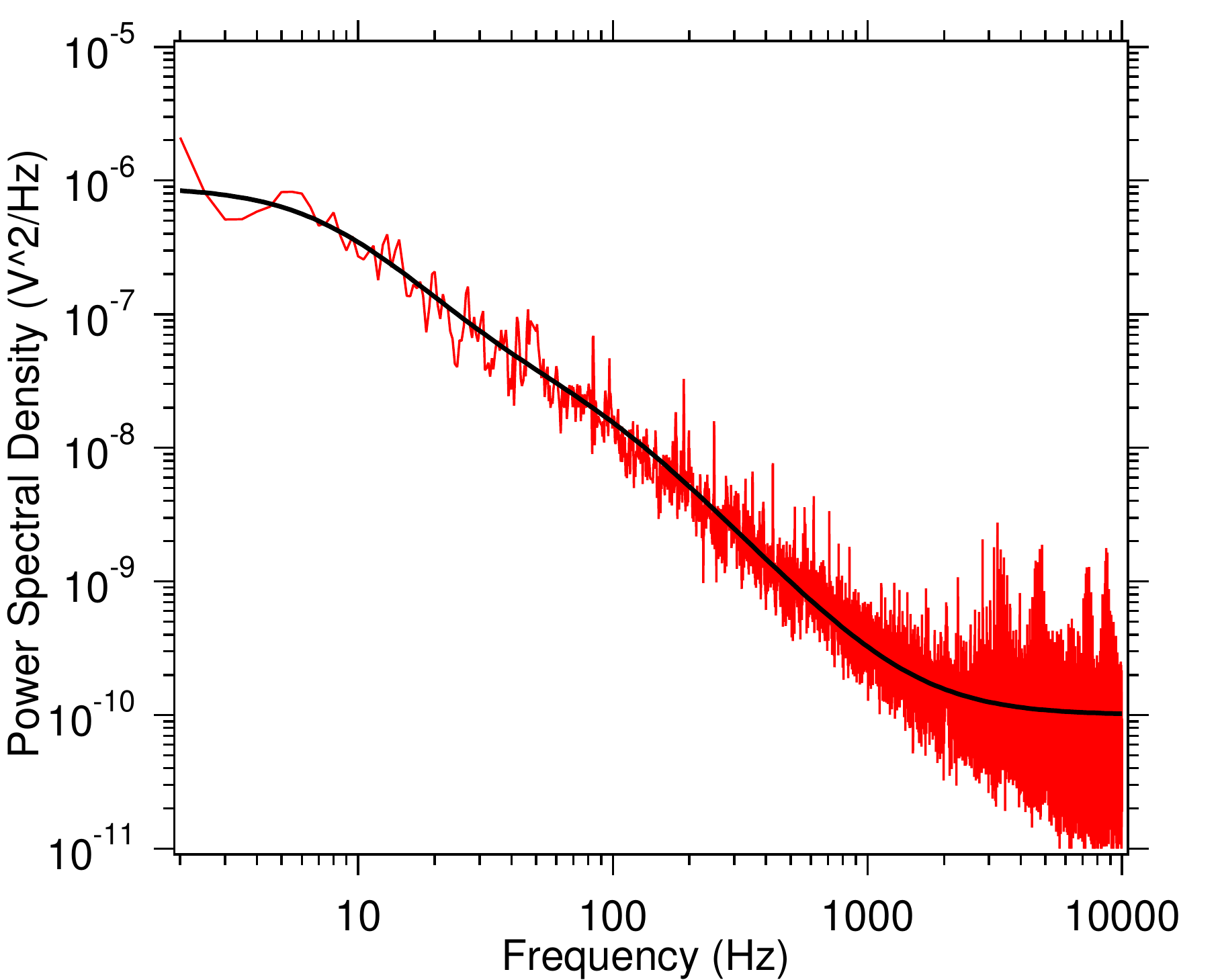}
 \caption{ This figure demonstrates a typical fit using eq. (\ref{fit}) to the case of a viscoelastic solution made from PAM in water. }
 \label{polymer}
\end{figure}

We first tested that our technique works for a solution of polymers in water. For this, a concentrated solution of polyacrylamide (PAM, 1$\%$ by weight) was prepared in a water solvent with suspended 1 $\mu$m diameter polystyrene particles to make the viscoelastic solution \cite{shuvojit2}. One such particle was trapped, and the power spectral density of translational motion recorded with the optical tweezers system. The curve has been mentioned in Fig. \ref{polymer} and fits well to the x-direction power spectral density dataset. 

In order to create in-vivo conditions, MCF-7 (Michigan Cancer Foundation-7) cells were grown on glass slides coated with gelatin. These glass slides were first treated with piranha solution and sterilized using UV (265nm) light for 20 minutes and coated with $ 0.5\% $ gelatin solution. MCF7 cells were added on to the center of the coverslip and Dulbecco's Modified Eagle Medium (DMEM) supplemented with 10$\%$ fetal bovine serum and 1$\%$ glutamine-penicillin-streptomycin was added on top of the coverslip. 10$\mu$L of gelatin-coated 0.5$\mu$m radius polystyrene particles($1\mu$g/mL) suspended in sterile serum-free media were sonicated and added to the cells. Cells were incubated with 
5\% carbondioxide and 37$^{\circ}$C for 12 hours to initiate endocytosis \cite{guo}.

 The temperature of the system was maintained using an air conditioner in the room at 26$^{\circ}$C. Care was taken to ensure that the cool air from the air-conditioner was not blowing directly onto the system. 

\begin{figure}
 \centering
 \includegraphics[width=\linewidth]{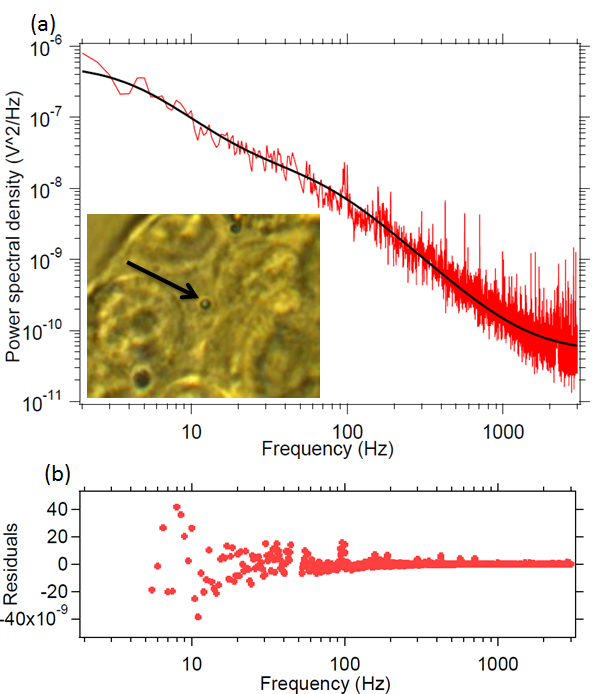}
 \caption{This figure represents the power spectral density (a) of x-displacements of a tracer particle optically trapped inside the cell. The data is well fitted by a viscoelastic equation of the form eq. \ref{fit}. The inset shows an image of the tracer particle of 1 $\mu$m diameter trapped inside the cell. (b) The residuals to the fit.}
 \label{invivo}
\end{figure}

The particles incubated with the cells might either be outside the cell or inside. We ascertain this by first trapping the particle and then moving the stage back and forth manually in the X direction using micrometer screws by about 10 $\mu$m. If the particle is outside the cell and not residing on the membrane, it shall readily follow the trap without too much delay. If the particle is residing on the membrane and also non-specifically bound, it cannot follow the trap much. However, if the particle is indeed inside the cell, it follows the trap within the space available to it without hitting the side membrane or organelles like the nucleus. One such video has been shown in S1. Once we ascertain the focus of the objective for which the particles can be trapped inside the cell, we mark it to find other particles at the same depth. 

\section{Results and discussions}

\begin{figure}
    \centering
    \includegraphics[width=\linewidth]{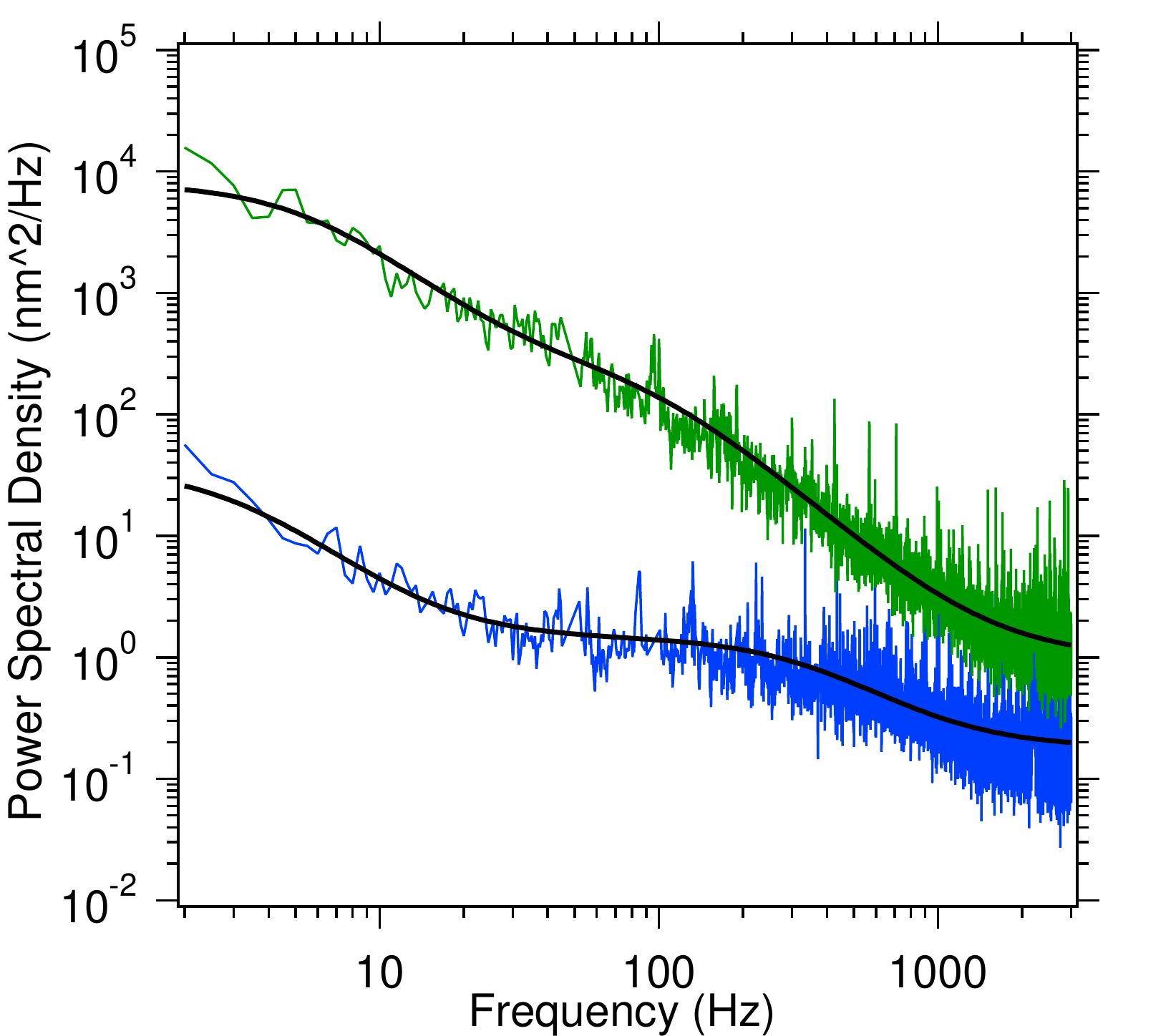}
    \caption{Power spectral density of the trapped particle fitted with the Oldroyd model, at different regimes of the same cell. The green curve indicates a typical PSD with DC viscosity about 5 times that of solvent while the blue curve indicates another location where it is 13 times that of solvent.  We show that there are regions showing higher DC viscosity inside the same cell by more than a factor of 2.}
    \label{sample}
\end{figure}

A typical motional power spectral density for a particle inside a cell has been shown in Fig. \ref{invivo}. Every power spectra is calculated by taking a 5 second time series and averaging over 10 such spectra. The eq. (\ref{fit}) fits well to the experimental data. The fitting parameters automatically yield the A parameter in Volts$^2$/Hz and subsequently the calibration factor $\beta$, as shown in eq, (\ref{fit2}). We also extract the ratios $\frac{\kappa}{\gamma_0}$ and $1+\frac{\mu_p}{\mu_s}$. We show two more datasets indicating PSD's at two different locations of the same cell in fig. \ref{sample}. The green curve shows a typical DC viscosity of 5 times that of solvent while the blue curve shows a value of 13 times that of solvent. There can be variations by  factors of 2 to 3 inside the same cell. These spectra are also calibrated to indicate thermal motion at each frequency. We can see that the DC thermal motion is of the order of 100 nm which reduces to 3 nm at 1000 Hz. Thus, we typically achieve a frequency range of about 3000 Hz with our present configuration. 

\begin{figure}
    \centering
    \includegraphics[width=\linewidth]{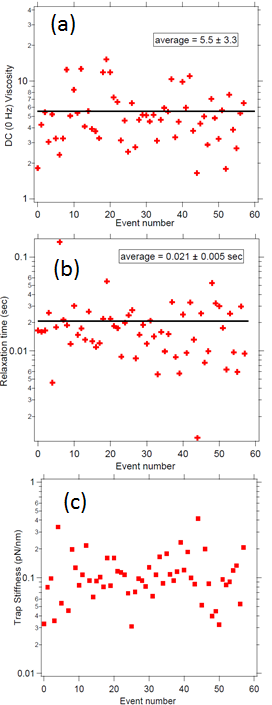}
    \caption{(a) The value of DC(0Hz) viscosity for the cell cytoplasm for 58 number of events.   (b)Measurement of polymer time constant $\lambda$ for the passive motion of the probe particle in every 58 events. (c) This shows  typical trap stiffness of 1 $\mu$m diameter polystyrene particles in-vivo. }
    \label{collated}
\end{figure}

 If we assume that the solvent medium for the cell is water, which has been proved to be a good approximation \cite{solvent}, the $\mu_s$ is automatically that of water while the $\gamma_0$ is given by eq. (\ref{solvent}). Using this approximation, we extract values for the DC viscosity $\mu_0 + \mu_p$, indicated in fig. \ref{collated}(a) and $\lambda$, indicated in fig. \ref{collated}(b) for 58 events.  

The DC (0 Hz) viscosity for the cell cytoplasm seems to vary from 2 mPa sec to 16 mPa sec, which is what one expects for typical intracellular concentrations of protein filaments like actin, microtubules and intermediate filaments dissolved in water\cite{proteinconc}. Further, the average relaxation time is 0.021 $\pm$ 0.005 sec. A typical estimate using the Maxwell's model observed a value of 0.2 sec\cite{magnetic}.

We also extract the typical trap stiffness at the same value of laser power of 200 mW at the sample plane, shown in fig. \ref{collated}(c). The trap stiffness seems to have an average value of 0.1 pN/nm. 

In these measurements, we span a frequency range between 2 Hz and 3000 Hz, partly including the athermal fluctuations inside the cell ranging from 0.1 Hz to 10 Hz\cite{schmidt}. Indeed we find deviations from the thermal curve below 5 Hz, as shown in Figs. \ref{invivo} and \ref{sample}. Moreover, as indicated by Tassieri \cite{tassieri}, the tweezers can indeed probe the viscoelasticity of the cell if the measurement time is lower than active motion time, given by the Deborah number (De) being greater than 1. Since we have a frequency range extending to 3 KHz, the measurmeent time is indeed smaller than 5 Hz activity time, thus enabling this kind of approach. Once the optical tweezers is calibrated properly, newer experiments could even be designed to look specifically at the athermal fluctuations.  

\begin{figure}
    \centering
    \includegraphics[width=\linewidth]{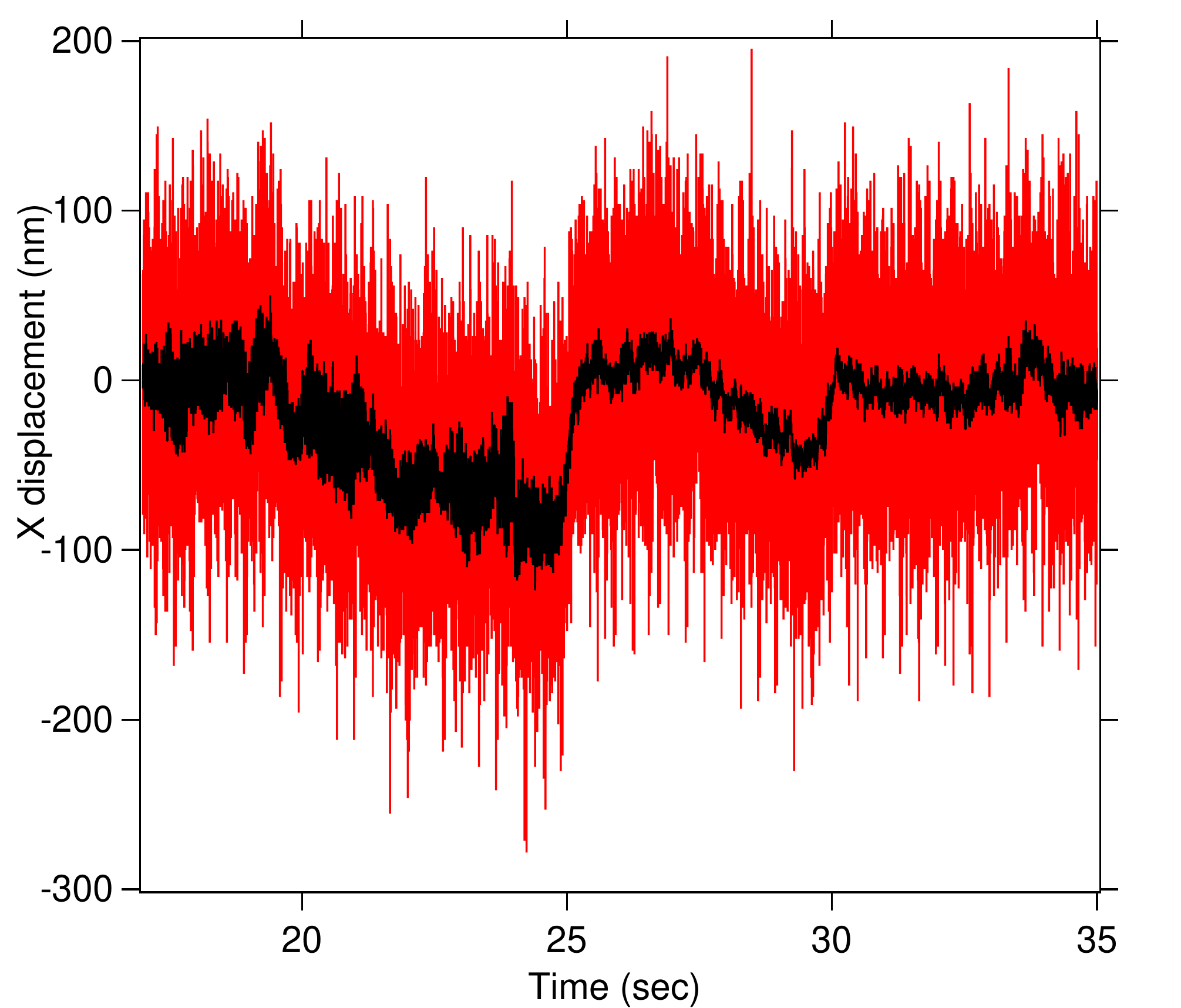}
    \caption{This figure indicates a calibrated x-position time series of a 1 $\mu$m diameter particle moving on a microtubule inside the cell. The red line indicates unfiltered data while the black curve indicated median filtered data.}
    \label{time_series}
\end{figure}

We also calibrate a typical time series for the x displacement of a 1 $\mu$m diameter microsphere inside the cell, while being carried by molecular motors on microtubules. It has been shown in Fig. \ref{time_series}. The red curve shows the unfiltered data while the black curve shows a median filtered data. The optical trap stiffness in this case was 0.12 pN/nm. Thus the force applied by the molecular motors on the particle is then about 10 pN, possibly indicating multiple motors at work. 

\section{Conclusions}
In conclusion, we could trap 1$\mu$m diameter polystyrene particle inside MCF-7 cells and obtain the power spectral density for passive motion of the particle along the X-axis inside the cell. To fit the PSD, we model the cytoplasm as a polymer network immersed in water that acts as a viscoelastic medium. This power spectrum is fitted with Jefferey's model. We see that the relative viscosity, polymer relaxation time and the trap stiffness varies inside the cell from place to place possibly due to the variation of cytoplasmic density, but maintains a good agreement with previous literature. This study of the viscoelasticity of the cell is made possible by the enhanced frequency response due to the power spectral density extending to 3 KHz, thereby ensuring a Deborah number greater than 1, when compared with activity frequency of 5 Hz or slower.

\section{Author contributions}

B.R. and S.B. designed the experiment. R. V., S.R.V. and B.R. performed the experiment and did data analysis. B. R. and S. B. wrote the manuscript. 

\section*{Acknowledgements}
We thank the Indian Institute of Technology Madras, India for their seed grant. 





\bibliography{rsc} 
\bibliographystyle{rsc} 

\end{document}